\documentstyle[preprint,floats,prl,aps,epsf]{revtex}

\begin{document}

\preprint{\hbox {June 2003}}

\draft
\title{Consistency Conditions for AdS/CFT Embeddings}
\author{\bf Paul H. Frampton$^{(a)}$ and Thomas W. Kephart$^{(b)}$}
\address{(a)Department of Physics and Astronomy,\\
University of North Carolina, Chapel Hill, NC  27599.}
\address{(b)Department of Physics and Astronomy,\\
 Vanderbilt University, Nashville, TN 37325.}
\maketitle
\date{\today}

\begin{abstract}
The group embeddings used in orbifolding the AdS/CFT correspondence
to arrive at quiver gauge field theories
are studied for both supersymmetric and non-supersymmetric
cases. For an orbifold $AdS_5 \times S^5/\Gamma$ the
conditions for embeddings of the finite group
$\Gamma$ in the $SU(4) \sim O(6)$ isotropy of $S^5$
are stated in the form of consistency rules, both
for Abelian and Non-Abelian $\Gamma$.

\end{abstract}

\newpage

\noindent {\it Introduction}

\bigskip

\noindent Independently of the important question of whether or not
superstring theory, or M theory, is
the correct theory of quantum gravity, superstring theory
can be used through the AdS/CFT correspondence\cite{MWGKP}
to arrive at interesting gauge field
theories in four spacetime dimensions.
According to the AdS/CFT correspondence, a Type IIB superstring compactified
on an $AdS_5 \times S^5$ background is dual to
an ${\cal N} = 4$ supersymmetric gauge field theory
in flat 4-dimensional spacetime.
The ${\cal N} = 4$ theory is much too symmetric to be
of phenomenological interest. What makes the AdS/CFT
correspondence far more interesting is the possibility
of orbifolding $S^5 \rightarrow S^5/\Gamma$ where $\Gamma$ is a finite group
embedded in the isometry
$SU(4) \sim O(6)$ of $S^5$ in the 
3-dimensional complex space ${\cal C}_3 \sim {\cal R}_6$.
Although $O(6)$ and $SU(4)$ have the same local structure, {\it i.e.} Lie algebra,
globally $O(6)$ is doubly-covered by $SU(4)$ and this distinction
will be important to us here.

Another consideration, already discussed fully in \cite{FK},
is the survival of chiral fermions. The correct statement
of the necessary and sufficient condition
is that the {\bf 4} of $SU(4)$
must be neither real nor pseudoreal for chiral fermions to be
present in the quiver gauge field theory.

It is often stated that the number of surviving
supersymmetries is ${\cal N} = 2, 1, 0$ according
to whether $\Gamma \subset SU(2), SU(3), SU(4)$.
While this statement essentially survives our
analysis, the statement itself needs sharpening,
in particular by a more careful definition
of the embedding. The present article
is devoted to addressing that issue.

\bigskip
\bigskip

\noindent {\it ${\cal N} = 1$ Embeddings of Abelian $\Gamma$ in SU(4) and O(6)}.

\bigskip

\noindent To preserve ${\cal N} = 1$ supersymmetry one must keep exactly one
invariant spinor under the joint action of the finite
symmetry $\Gamma$ and the
quiver gauge group, which for Abelian $\Gamma$ is $SU(N)^p$ where
$p$ is the order of $\Gamma$. This implies that one component of the {\bf 4}
of $SU(4)$ is the trivial singlet representation of $\Gamma$.

The most general Abelian $\Gamma$ of order $g=o(\Gamma)$ is \cite{books,FK2} 
made up of
the basic units $Z_p$, the order p cyclic groups formed from the $p^{th}$ roots
of unity. It is important to note that the the product $Z_pZ_q$ is identical
to $Z_{pq}$ if and only if p and q have no common prime factor.

If we write the prime factorization of g as:
\begin{equation}
g = \prod_{i}p_i^{k_i}
\end{equation}
where the product is over primes, it follows that the number
$N_a(g)$ of inequivalent abelian groups of order g is given by:
\begin{equation}
N_a(g) = \prod_{k_i}P(k_i)
\end{equation}
where $P(x)$ is the number of unordered partitions of $x$.
For example, for order $g = 144 = 2^43^2$ the value would be
$N_a(144) = P(4)P(2) = 5\times2 = 10$. For $g\leq31$ it is simple
to evaluate $N_a(g)$ by inspection. $N_a(g) = 1$ unless g contains
a nontrivial power ($k_i\geq2$) of a prime. These exceptions are:
$N_a(g = 4,9,12,18,20,25,28) = 2; N_a(8,24,27) = 3$; and $N_a(16) = 5$.
This confirms that:
\begin{equation}
\sum_{g = 1}^{31}N_a(g) = 48
\end{equation}

Let us define $\alpha = exp(2 \pi i / p)$ and write the
embedding as ${\bf 4} = ( \alpha^{A_1}, \alpha^{A_2}, \alpha^{A_3},
\alpha^{A_4})$. For ${\cal N} = 1$ we have agreed that one $A_{\mu}$ must vanish
so let us put $A_4 = 0$
and denote the embedding by the shorthand $\underline{{\bf A}} \equiv (A_1, A_2, A_3)$.

Consider now the {\bf 6} of $SU(4)$ which is the antisymmetric
part in $({\bf 4} \times {\bf 4})$. The {\bf 6}
is necessarily real, and in the double covering of $O(6)$
is identified with the defining "vector"
representation of $O(6)$. In the shorthand notation
already adopted for the {\bf 4},
the {\bf 6} is
${\bf 6} = (A_1, A_2, A_3, A_1+A_2, A_2+A_3, A_3+A_1)$.

For general $\underline{{\bf A}}$, the {\bf 6} is not real
but we must now impose the condition that the
diagonal $3 \times 3$ matrix with diagonal elements 
$\alpha^{\underline{{\bf A}}}$ 
be a transformation belonging to $SU(3)$,
{\it i.e.} a matrix which is unitary and of determinant one.
This requires that $\sum_i A_{i} = 0$ (mod p).
But with that condition we can rewrite (up to mod p)
${\bf 6} \equiv (A_1, A_2, A_3, -A_3, -A_2, -A_3)$
which is manifestly real and hence we have a consistent embedding.

Thus, in this the simplest case, the embedding is uniquely
defined for the 
{\bf 4} of $SU(4)$ by the three components (recall $A_4 \equiv 0$)
of $\underline{{\bf A}} \equiv (A_1, A_2, A_3)$ which
for consistency must satisfy 
$A_1+A_2+A_3 = 0$ (mod p).
This will always lead to a consistent ${\cal N} = 1$
chiral quiver gauge theory.

\bigskip
\bigskip

\noindent {\it ${\cal N} = 0$ Embeddings of Abelian $\Gamma$ in SU(4) and O(6)}.

\bigskip

\noindent For ${\cal N} = 0$ the discussion is similar to the above
except that $A_4$ does not vanish. Therefore the considerations
involve not $\underline{{\bf A}}$ with three components but
$A_{\mu} = (A_1, A_2, A_3, A_4)$ with four components.

Consider now the {\bf 6} of $SU(4)$ which is the antisymmetric
part in $({\bf 4} \times {\bf 4})$.
The {\bf 6} can be written 
${\bf 6} = (A_1+A_4, A_2+A_4, A_3+A_4, A_1+A_2, A_2+A_3, A_3+A_1)$.

But here for consistency the $4 \times 4$ diagonal matrix
with diagonal elements $\alpha^{A_{\mu}}$ must be an $SU(4)$
transformation which requires $\sum_{\mu} A_{\mu} = 0$ (mod p)
in which case the {\bf 6} is again manifestly real.

For a consistent ${\cal N} = 0$ quiver gauge field theory, therefore,
the {\bf 4} of $SU(4)$ is defined by four non-vanishing
integers $A_{\mu}$ satisfying $A_1+A_2+A_3+A_4 = 0$ (mod p). 
This is the necessary and sufficient condition for consistent
Abelian orbifolding to ${\cal N} = 0$. Similar results hold for products of abelian groups, e.g.,
for $\Gamma=Z_pZ_Q$ we can write $\bf{4}=(e^{A_1}e^{B_1},e^{A_2}e^{B_2},e^{A_3}e^{B_3},e^{A_4}e^{B_4})$
where we require $\sum_{\mu} A_{\mu} = 0$ (mod p) and $\sum_{\mu} B_{\mu} = 0$ (mod q).

\bigskip
\bigskip

\noindent {\it ${\cal N} = 1$ Embeddings of Non-Abelian $\Gamma$ in SU(4) and O(6)}.

\bigskip

As usual, going from Abelian $\Gamma$ to Non-Abelian $\Gamma$ makes
everything more complicated. This is why paper \cite{FK} is necessarily much
longer than {\it e.g.} paper \cite{PHF}!

The complication arises because for Non-Abelian $\Gamma$, the
irreducible representations $\rho_i$ can have dimensions $d_i$
greater than one. If the order of $\Gamma$ is $g$ the dimensions
must satisfy $\Sigma_i d_i^2 = g$. The resultant
gauge group for the quiver theory is $\bigotimes_i SU(nd_i)$.
Because of this more complex quiver theories emerge
from Non-Abelian $\Gamma$ than are possible for Abelian $\Gamma$.
On the other hand, as shown in \cite{FK}, the constraints
imposed phenomenologically are also more complex to such an extent
that for all $g \leq 31$ only one acceptable theory occurs
amongst many hundreds of candidates. 

For the Non-Abelian embeddings we must again study
whether the ${\bf 6} \equiv ({\bf 4} \times {\bf 4})_a$ is consistently real.
Let us, for the ${\cal N} = 1$ case write ${\bf 4} = (\sum_{i^{'}} \rho_{i^{'}}, 1)$
where $\sum d_{i^{'}} = 3$ and where none of the $\rho_{i^{'}}$ is the
identity representation. 
(Here the set \{$i^{'}$\} is a subset of the \{$i$\}). 
This will generically lead to an ${\cal N} = 1$
supersymmetric quiver gauge theory.

The consistency requirements for this Non-Abelian embedding are
very demanding and may be stated as the three rules for ${\cal N} = 1$:

\bigskip

\noindent 1. Exactly one component of the {\bf 4}
of $SU(4)$ must be the identity of $\Gamma$.
That is, 

\noindent {\bf 4} = ($\sum_{i^{'}} \rho_{i^{'}}$, 1) where the $\rho_{i^{'}}$ 
are irreducible representations
(not the identity) of $\Gamma$ satisfying $\sum d_{i^{'}} = 3$.

\bigskip

\noindent 2. Written as a block-diagonal matrix the determinant of the $3 \times 3$
matrix $\otimes_{i^{'}} \rho_{i^{'}}$ must be real and equal to one. Note that
this reduces to the vanishing trace condition if we were to break
$\Gamma$ to an abelian subgroup.

\bigskip

To see just how demanding these three rules are consider even the simplest
Non-Abelian finite group $\Gamma = S_3 \equiv D_3$. The order is g = 6 and the
available irreducible representations are $\rho = 1, 1', 2$. 
The only choice satisfying Rules 1 and 2 is 
{\bf 4} = $(1^{'}, 2, 1)$ and this satisfies Rule 2 because 
$2$ is an $SU(2)$ matrix which
necessarily has determinant equals one.

\bigskip
\bigskip

\newpage

\noindent {\it ${\cal N} = 0$ Embeddings of Non-Abelian $\Gamma$ in SU(4) and O(6)}.

\bigskip

The consistency requirements for Non-Abelian embedding to
give ${\cal N} = 0$ are similar to but different from those 
given for ${\cal N} = 0$ and may be stated as the two new rules:

\bigskip

\noindent 1. No component of the {\bf 4}
of $SU(4)$ may be the identity of $\Gamma$.
That is, 

\noindent {\bf 4} = ($\sum_{i^{'}} \rho_{i^{'}}$) where the $\rho_{i^{'}}$ 
are irreducible representations
(not the identity) of $\Gamma$ satisfying $\sum d_{i^{'}} = 4$.

\bigskip

\noindent 2. Written as a block-diagonal matrix the determinant of the $4 \times 4$
matrix $\otimes_{i^{'}} \rho_{i^{'}}$ must be real and equal to one.

\bigskip

To check how these rules work, consider again the simplest
Non-Abelian finite group $\Gamma = S_3 \equiv D_3$. 
The only choice satisfying Rules 1 and 2 is 
{\bf 4} = (2, 2).
Therefore $\Gamma = S_3$ can be used to obtain 
either an ${\cal N} = 1$ or an ${\cal N} = 0$
quiver gauge field theory.
[The only other consistent embeddings of $S_3$ in $SU(4)$
are {\bf 4} = $(1, 1, 1^{'}, 1^{'})$ which leads to
${\cal N} = 2$ and the trivial embedding which leaves 
${\cal N} = 4$.]

\bigskip
\bigskip
\bigskip

\newpage

\bigskip

\noindent {\it Summary and Discussion}.

\bigskip

Let us summarize the consistency conditions for the four cases considered:

{\bf Abelian $\Gamma$; ${\cal N} = 1$}.

$A_4 = 0$ (mod p) and $A_1+A_2+A_3 = 0$ (mod p).

{\bf Abelian $\Gamma$; ${\cal N} = 0$}

$A_1+A_2+A_3+A_4 = 0$ (mod p).

{\bf Non-Abelian $\Gamma$; ${\cal N} = 1$} 

{\bf 1.} Exactly one component of the {\bf 4}
of $SU(4)$ must be the identity of $\Gamma$.

{\bf 2.} The determinant of the $3 \times 3$
matrix $\otimes_{i^{'}} \rho_{i^{'}}$ must be real and equal to one.

{\bf Non-Abelian $\Gamma$; ${\cal N} = 0$}

{\bf 1.} No component of the {\bf 4}
of $SU(4)$ may be the identity of $\Gamma$.

{\bf 2.} The determinant of the $4 \times 4$
matrix $\otimes_{i^{'}} \rho_{i^{'}}$ must be real and equal to one.

\bigskip

Fortunately, these consistency conditions are respected
by most of the earlier work
on Abelian $\Gamma$. For example, the TeV unificiation
model in \cite{PHF2} which provides an alternative
to SUSYGUTs\cite{ADFFL} is consistent. Most of the
Abelian models in \cite{KP} are consistent but
7 of the 60 which are called non-partition are not.
Even though the embeddings of these seven models are inconsistent, the
models are themselves consistent gauge theories and of interest in their
own right. Despite the fact that they are not derivable from orbifolding
$AdS\times S^{5}$, they have many of the features of orbifolded
$AdS\times S^{5}$
models, $e.g$., vanishing first-order (and perhaps higher order) $\beta $
functions, similar boson-fermion counting providing ${\cal N}=1$ SUSY, etc.
It can be shown that these are the only models besides the partition models
with a real ${\bf 6}$, and that they fall into three classes. Recall first
that the ${\cal N}=1$ partition models for $\Gamma =Z_{p}$ can be
written as
$M_{p_{1},p_{2},p_{3}}^{p}$ where $p_{1}+p_{2}+p_{3}=0 (mod p)$. In this
notation the nonpartition models are $M_{p,p,2p}^{3p},$
$M_{p,2p,4p}^{6p},$ and $M_{p,4p,7p}^{9p}$. Along with their own
phenomenological interest, these models can probably be used in comparison
with properly embedded partition orbifolded models to investigate violation
of conformal invariance, etc. In the ${\cal N} = 0$ case, again one finds
partition models $M_{p_{1},p_{2},p_{3},p_{4}}^{p}$
with $p_{1}+p_{2}+p_{3}+p_{4}=0 (mod p)$ are properly embedded.
But again there exist improperly-embedded, but phenomenologically-interesting,
non-partition models with complex ${\bf 4}$ and real
${\bf 6}$\cite{KP2}.

For Non-Abelian $\Gamma$ the most extensive
work is in our earlier paper \cite{FK}. 
The only phenomenologically-acceptable model out of the many considered in
\cite{FK2} was for the choice $\Gamma \equiv 24/7$,
in the notation of Thomas and Wood\cite{books} for $D_4 \times Z_3$.
The embedding used is ${\bf 4} = (1\alpha, 1^{'}, 2\alpha)$
and this satisfies Rules 1 and 2 for Non-Abelian $\Gamma$ and 
${\cal N} = 0$. It is therefore a quiver gauge theory
which arises from a consistent embedding of $\Gamma$ in
$AdS_5 \times S^5/\Gamma$. 

\bigskip
\bigskip

\newpage

\noindent {\it Acknowledgements}

\bigskip

\noindent TWK thanks the Department of Physics and Astronomy at UNC Chapel Hill 
and PHF thanks the Department of Physics and Astronomy
at Vanderbilt University for hospitality
while this work was in progress. This work was supported in part by the
US Department of Energy under
Grants No. DE-FG02-97ER-41036 and No. DE-FG05-85ER40226.

\bigskip
\bigskip
\bigskip

\end{document}